\documentclass[sigconf,natbib=true,screen=true]{acmart}

\author{Philipp Hager}
\orcid{0000-0001-5696-9732}
\authornote{Both authors contributed equally to the paper}
\affiliation{%
    \institution{University of Amsterdam}
    \city{Amsterdam}
    \country{The Netherlands}
}
\email{p.k.hager@uva.nl}

\author{Romain Deffayet}
\orcid{0000-0001-8265-9092}
\authornotemark[1]
\affiliation{%
    \institution{University of Amsterdam}
    \city{Amsterdam}
    \country{The Netherlands}
}
\email{romain.deffayet@naverlabs.com}

\author{Jean-Michel Renders}
\orcid{0000-0002-7516-3707}
\affiliation{%
    \institution{Naver Labs Europe}
    \city{Meylan}
    \country{France}
}
\email{jean-michel.renders@naverlabs.com}

\author{Onno Zoeter}
\orcid{0000-0003-1704-706X}
\affiliation{%
    \institution{Booking.com}
    \city{Amsterdam}
    \country{The Netherlands}
}
\email{onno.zoeter@booking.com}

\author{Maarten de Rijke}
\orcid{0000-0002-1086-0202}
\affiliation{%
    \institution{University of Amsterdam}
    \city{Amsterdam}
    \country{The Netherlands}
}
\email{m.derijke@uva.nl}

\usepackage[inline]{enumitem}
\usepackage{multirow}
\usepackage{acronym}
\usepackage{stfloats}
\usepackage[skip=0pt]{caption}
\usepackage{etoolbox}

\makeatletter 
\newcommand\mynobreakpar{\par\nobreak\@afterheading} 
\makeatother

\acrodef{LTR}{learning-to-rank}
\acrodef{ULTR}{unbiased learning-to-rank}

\acrodef{PBM}{position-based model}
\acrodef{UBM}{user browsing model}
\acrodef{IPS}{inverse propensity scoring}
\acrodef{DLA}{dual learning algorithm}
\acrodef{RegressionEM}{regression expectation-maximization}
\acrodef{PairD}{pairwise debiasing}

\acrodef{DCG}{discounted cumulative gain}
\acrodef{MRR}{mean reciprocal rank}
\acrodef{NLL}{negative log-likelihood}
\acrodef{CTR}{click-through rate}

\input{preamble/meta}

\begin{document}

\title[Unbiased Learning to Rank Meets Reality: Lessons from Baidu's Large-Scale Search Dataset]{Unbiased Learning to Rank Meets Reality:\\ Lessons from Baidu's Large-Scale Search Dataset}



\begin{abstract}
    Unbiased learning-to-rank (ULTR) is a well-established framework for learning from user clicks, which are often biased by the ranker collecting the data. While theoretically justified and extensively tested in simulation, ULTR techniques lack empirical validation, especially on modern search engines. The Baidu-ULTR dataset released for the WSDM Cup 2023, collected from Baidu's search engine, offers a rare opportunity to assess the real-world performance of prominent ULTR techniques. Despite multiple submissions during the WSDM Cup 2023 and the subsequent NTCIR ULTRE-2 task, it remains unclear whether the observed improvements stem from applying ULTR or other learning techniques.
    
    In this work, we revisit and extend the available experiments on the Baidu-ULTR dataset. We find that standard unbiased learning-to-rank techniques robustly improve click predictions but struggle to consistently improve ranking performance, especially considering the stark differences obtained by choice of ranking loss and query-document features. Our experiments reveal that gains in click prediction do not necessarily translate to enhanced ranking performance on expert relevance annotations, implying that conclusions strongly depend on how success is measured in this benchmark.
\end{abstract}

\maketitle

\vspace{-1mm}
\section{Introduction}

Historically, the field of \acl{LTR} employed human experts to annotate the relevance of search results to create training data for ranking models~\citep{Qin2013MSLR,Dato2016Istella,Dato2022Istella22}. As expert labeling is expensive~\cite{Chapelle2011Yahoo}, infeasible in certain applications~\citep{Wang2016IPW}, and potentially misaligned with user preferences~\citep{Sanderson2010TREC}, many practitioners seek to leverage implicit user feedback, often collected in the form of clicks. However, clicks are usually a biased signal of relevance, as they often depend on the position of an item~\citep{Joachims2005EyeTracking}, its surrounding items~\citep{Craswell2008PBM,Zhuang2021XPA}, or even the user's trust in the system to place relevant items on top~\citep{Joachims2005EyeTracking,Agarwal2019TrustBias,Vardasbi2020Affine}. 
Over the years, the field of \acf{ULTR} has proposed methods to mitigate biases when training ranking models on click~data~\cite{Gupta2024ULTR}.

\acs{ULTR} methods, especially in academic research, are often evaluated in synthetic or semi-synthetic setups~\citep{Joachims2017IPW,Ai2018DLA,Vardasbi2020Affine,Vardasbi2020Cascade,Oosterhuis2021LambdaLoss}. The most common semi-synthetic setup goes back to the seminal work by \citet{Joachims2017IPW}. The setup uses \acl{LTR} datasets from web search engines containing features and expert judgments of each query-document pair~\citep{Chapelle2011Yahoo,Qin2013MSLR,Dato2016Istella} and simulates clicks and biases based on synthetic user models (e.g., the \acf{PBM}~\citep{Richardson2007PBM,Craswell2008PBM} or the cascade model~\citep{Craswell2008PBM}).

\begin{figure}
    \centering
    \hspace*{-2em}
    \includegraphics[clip, trim=0mm -20mm 0mm 0mm, width=0.9\linewidth]{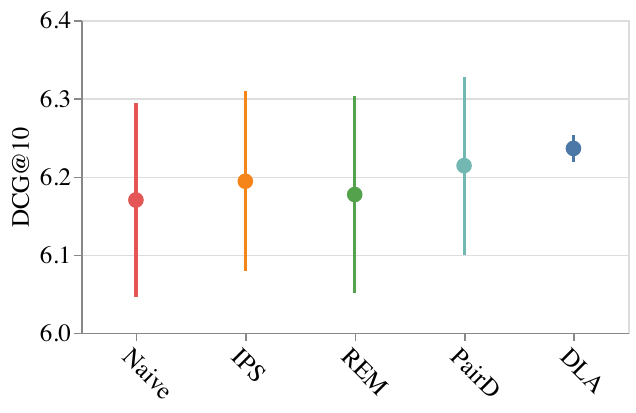}

    \vspace*{-1.4cm}
    \caption{The original experiments on Baidu-ULTR show that none of the four compared ULTR methods outperform a naive method not correcting for position bias. Data from \cite[Table 6]{Zou2022Baidu} and visualization by us.}
    \label{fig:baidu}
    \vspace*{-3mm}
\end{figure}

While the semi-synthetic simulation setup has demonstrated the effectiveness of many \acs{ULTR} methods, it is often unclear to academic researchers how these methods fare in a real-world setup. Recently, \citet{Zou2022Baidu} released Baidu-ULTR, an extensive real-world dataset for web search. It comprises over $1.2$ billion user sessions with click data and $397,572$ annotated query-document pairs for evaluation. The dataset contains rich user feedback, including clicks, dwell time, whether a document was scrolled off-screen, and whether the user returned to the result page after clicking an item. In contrast to the classic \acs{LTR} datasets used in semi-synthetic simulation setups, the dataset does not contain pre-computed ranking features, but the original search query, title, and abstract of each result (for privacy reasons in tokenized form, with a private vocabulary). Therefore, the dataset enables training and evaluating transformer-based rankers, such as MonoBERT~\cite{Nogueira2019MonoBERT} or MonoT5~\cite{Pradeep2021MonoT5}.

\citet{Zou2022Baidu} train a MonoBERT ranker and provide baseline results of this model fine-tuned on four standard \acs{ULTR} methods: \acf{IPS}~\cite{Wang2016IPW,Joachims2017IPW}, \acf{RegressionEM}~\cite{Wang2018RegressionEM}, the \acf{DLA}~\cite{Ai2018DLA}, and \acf{PairD}~\cite{Hu2019UnbiasedLambdaMART}. Interestingly, \citet{Zou2022Baidu} find that these prominent ULTR methods overall do not perform significantly better than a naive BERT model trained on clicks without any bias correction,\footnote{\citet{Zou2022Baidu} find that DLA outperforms a naive baseline on frequent head queries.} as we report in Figure~\ref{fig:baidu}.
In subsequent work, the Baidu-ULTR dataset was used at multiple ranking competitions (the WSDM Cup 2023~\citep{WSDMCup2023ULTR,WSDMCup2023Pretrain} and NTCIR's ULTRE-2~\citep{Niu2023-ULTRE2}), in which multiple teams reported noticeably higher ranking performance, with and without \acs{ULTR} techniques. \emph{It therefore remains unclear whether the observed improvements were due to \acl{ULTR}, and ultimately whether \acs{ULTR} techniques improve performance on this dataset.}

In this work, we reproduce and extend the experiments by \citet{Zou2022Baidu}, in light of several considerations:
\begin{enumerate}[nosep, leftmargin=*, label=(\arabic*)]
    \item The initial findings of \acs{ULTR} methods bringing barely any benefits on the largest real-world dataset available for \ac{ULTR} have substantial implications for the field and warrant more scrutiny.
    \item Three competitions have been conducted on the Baidu-ULTR dataset, with participants reporting vastly improved ranking results, though not necessarily using \acs{ULTR}. A preliminary experiment by us revealed that a random ranking of the annotated test dataset achieves a \acs{DCG}@10 $\approx 6.69$, which is higher than all baselines in Figure~\ref{fig:baidu} and suggests potential issues with the original experiments.
    \item According to the official codebase that comes with the Baidu-ULTR dataset,\footnote{\url{https://github.com/ChuXiaokai/baidu_ultr_dataset/}} the original comparison focuses on pointwise versions of the compared ranking methods (except for pairwise debiasing). Therefore, it remains open how pointwise, pairwise, and listwise \acs{ULTR} methods compare on this dataset.
    \item And lastly, a closer look at the Baidu-ULTR dataset and the original codebase revealed some disputable design decisions. \acs{IPS}, for example, requires an estimate of position bias on the given dataset, while the creators of the Baidu-ULTR dataset confirmed using a hardcoded position bias from the ULTRA library in their experiments~\citep{Tran2021ULTRA}. We also found that placeholder documents make up almost $20\%$ of the Baidu-ULTR dataset (see Section~\ref{sec:analysis}), with an unknown impact on the trained models.
\end{enumerate}

\noindent By revisiting the experiments reported by~\citet{Zou2022Baidu} and the cups' participants, we address the following research questions:
\begin{enumerate}[nosep, leftmargin=*, label = \textbf{(RQ\arabic*)}]
    \item Does \acl{ULTR} improve performance on the Baidu-ULTR dataset over naive, non-debiasing models?
    \item How do \acs{ULTR} methods fare against each other, and how do ranking losses and input features affect their performance?
    \item Can \acs{ULTR} methods be applied during language model training, and do they bring improved performance?
\end{enumerate}

\noindent%
We answer (RQ1) negatively as we cannot find robust improvements in ranking performance in a fair comparison between ULTR-based and naive methods. For (RQ2), we report minor but existing differences between ULTR methods and showcase large and reliable differences due to ranking losses and input features. Regarding (RQ3), we find that while certain ULTR methods improve over their naive counterpart if applied during language model training, the interactions of ULTR and transformer-based models are not well understood and can even lead to decreased performance.

Besides answering our research questions, our contributions are:\mynobreakpar
\begin{itemize}[nosep, leftmargin=*]
    \item We release two smaller, cleaned, and pre-processed datasets derived from Baidu-ULTR,\footnote{\url{https://huggingface.co/datasets/philipphager/baidu-ultr_uva-mlm-ctr}} along with transformer-based query-document embeddings to enhance the reproducibility of our work and ease access to Baidu-ULTR.
    \item We publish highly optimized Jax~\citep{Bradbury2018Jax,Jagerman2022Rax} implementations of various standard \acs{ULTR} methods to enable \acs{ULTR} researchers to leverage the vast scale of the Baidu-ULTR dataset effectively.\footnote{\url{https://github.com/philipphager/ultr-reproducibility}}
    \item We tune six \acs{ULTR} methods and three naive ranking methods on transformer-based embeddings and learning-to-rank features.
    \item We train six MonoBERT models from scratch, including listwise and \acs{ULTR}-based loss functions; our model weights are public.\footnote{\url{https://github.com/philipphager/baidu-bert-model}}
\end{itemize}

\section{Related Work}
\subsection{Unbiased Learning to Rank}

The field of \acl{ULTR} can be broadly divided into click modeling and counterfactual \acl{LTR}. Click modeling encodes assumptions on user behavior into probabilistic models~\citep{Craswell2008PBM,Dupret2008UBM,Chapelle2009DBM}. Assumed effects, such as position bias or item relevance, are represented as (latent) variables that are jointly inferred via maximum likelihood estimation~\citep{Chuklin2015ClickModels}. Notable examples include the \acfi{PBM}~\citep{Richardson2007PBM, Craswell2008PBM}, which assumes that users click only on positions they examine and items they find relevant, and the \emph{cascade model}~\citep{Craswell2008PBM}, which assumes that users scroll from top to bottom, click on the first relevant item, then leave the page. In recent years, more complex and flexible click models have been proposed using neural architectures, semantic embeddings, and bias features beyond position~\citep{Borisov2016NCM, Chen2020CACM, Zhuang2021XPA,Yan2022TwoTowers}. We implement two neural versions of the original PBM: RegressionEM~\citep{Wang2018RegressionEM} and an additive two-tower model popular in industry applications~\citep{Guo2019PAL,Zhang2023DisentanglingTwoTowers,Zhao2019AdditiveTowers}.

The counterfactual \acl{LTR} community has historically employed simpler user models such as the \acs{PBM}~\cite{Wang2016IPW,Joachims2017IPW} or the affine model for trust bias~\citep{Vardasbi2020Affine} and focused on training more effective ranking models, mitigating biases with inverse propensity scoring~\citep{Wang2016IPW,Joachims2017IPW,Oosterhuis2021LambdaLoss}. Our work includes an extended binary cross-entropy loss~\cite{Bekker2019PointwiseIPS, Saito2020PointwiseIPS} as a pointwise IPS baseline and an extended softmax loss as a listwise IPS baseline~\citep{Bruch2019Softmax,Ai2018DLA}. Both methods require position bias estimations, for which we release the implementation of three intervention harvesting methods (more in Section~\ref{sec:position_bias}). We also include two counterfactual methods jointly estimating position bias and item relevance, the \acf{DLA}~\citep{Ai2018DLA} and \acf{PairD}~\citep{Hu2019UnbiasedLambdaMART}. We introduce all unbiased learning to rank methods used in this work in Section~\ref{sec:method}.

\subsection{ULTR on the Baidu-ULTR dataset}
Most prior work on the Baidu-ULTR dataset has revolved around three public competitions: Two tracks at the WDSM Cup 2023 --\emph{\acl{ULTR}} \citep{WSDMCup2023ULTR} and \emph{pretraining for web search} \citep{WSDMCup2023Pretrain} -- and the ULTRE-2 track at NTCIR~\citep{Niu2023-ULTRE2}. 

In the \acs{ULTR} track of the WSDM cup, participants were restricted to train ranking models using click data. Two of the top three teams ended up applying \acs{ULTR} techniques~\cite{Chen2023-TENCENT-ULTR-1,Yu2023-CIR-ULTR-3}, notably the winning team which employed a two-tower model with a softmax loss~\citep{Chen2023-TENCENT-ULTR-1}. However, participants did not compare with non-\acs{ULTR} methods (as the main objective is to win a competition). All three top-performing teams incorporated a BERT-based cross-encoder~\citep{Nogueira2019MonoBERT}, with the first two teams training models from scratch~\citep{Chen2023-TENCENT-ULTR-1,Chen2023-THUIR-ULTR-2}. \citet{Chen2023-THUIR-ULTR-2} voice concerns about the performance of their BERT model trained using the officially released starter-kit, as it plateaued around a $\text{DCG@10}\approx7$. Their remark has led us to train our own BERT cross-encoder models from scratch. 

The second task of the WSDM Cup allowed participants to pre-train language models from clicks before fine-tuning on expert annotations. All top teams combined the output of their BERT models with traditional LTR features and tuned gradient-boosting models on annotations~\cite{Li2023-TENCENT-PRETRAIN-1,Li2023-THUID-PRETRAIN-2,Sun2023-CIR-PRETRAIN-3}. Only the third-placed team incorporated a conventional \acs{ULTR} approach with a softmax ranking loss and \acs{IPS} during BERT pretraining. 

In the NTCIR 17 ULTRE-2 track~\cite{Niu2023-ULTRE2}, the organizers released a subset of the data ($\approx 1$ million sessions) to make Baidu-ULTR more accessible, using the best-performing BERT model trained by \citet{Li2023-TENCENT-PRETRAIN-1} during the WSDM Cup to create query-document embeddings in combination with traditional lexical matching features.\footnote{Note that the first ULTRE-1 task at NTCIR 16 was not yet conducted on Baidu-ULTR but on a semi-synthetic simulation setup~\citep{Zhao2022-ULTRE1}.} In their baseline experiments, they find that \acs{DLA} models perform better than non-\acs{ULTR} models trained with a pointwise loss but only marginally better than those trained with a listwise loss. The only participating team, \citet{Yu2023-ULTRE2-CIR-1}, proposed to alleviate selection bias on items that are relevant but never clicked by using a \acs{DLA} model to re-annotate non-clicked documents. However, they do not compare with other \acs{ULTR} and non-\acs{ULTR} baselines.

Overall, participants in the three cups reported noticeably higher ranking performance than the original experiments by the Baidu-ULTR dataset authors~\citep{Zou2022Baidu}, and identified helpful techniques, such as re-weighting queries to address the long-tail query distribution~\cite{Sun2023-CIR-PRETRAIN-3}, negative sampling of documents~\cite{Sun2023-CIR-PRETRAIN-3}, ignoring clicks on items displayed for a short time~\cite{Li2023-TENCENT-PRETRAIN-1}, or pseudo-relevance feedback~\citep{Yu2023-ULTRE2-CIR-1}.
However, it remains unclear whether observed improvements stemmed from debiasing the click feedback or such orthogonal techniques.
In order to identify the contribution of click debiasing to the results, we extend the work by \citet{Niu2023-ULTRE2} and \citet{Zou2022Baidu} to larger datasets, more \acs{ULTR} methods, fairer baselines, and investigate the interactions of ULTR with language model training. Next, we introduce the dataset used in this work in more detail.

\section{Overview of Baidu-ULTR}
\label{sec:analysis}

\subsection{Description of the data}
\label{sec:description}
The Baidu-ULTR dataset~\citep{Zou2022Baidu} contains more than $1.2$ billion search sessions (split into $2,000$ dataset partitions) with user clicks and $7,008$ annotated queries for evaluation. The dataset was collected in April 2022 by randomly sampling the search traffic of Baidu~\citep{Zou2022Baidu}, reflecting the long-tail query distribution of real user traffic. The dataset contains logged queries, document titles, and abstracts of the search results presented to the user. The authors tokenized all released text with a private vocabulary for user privacy. In addition to tokenized text, the dataset also contains a multitude of logged user interactions, including clicks, dwell-time, and skipping behavior, as well as item presentation features, including document height, item type, and ranking position. In this work, we solely focus on item position and user clicks.

To foster the reproducibility of this work and to ease access to the vast Baidu-ULTR dataset for academic research, we release two smaller datasets: A \emph{language modeling dataset}, comprised of the first 125 partitions of the dataset for training transformer models from scratch, and a smaller \emph{reranking dataset} comprised of four partitions to compare \acs{ULTR} on pre-computed query-document features. In the following, we describe the preprocessing common to both datasets and analyze the properties of our reranking dataset.

\vspace{-1mm}
\subsubsection{Pre-processing}
\begin{enumerate*}[nosep, leftmargin=*, label=(\roman*)]
    \item We use the md5 hash of the query tokens and document URL, respectively, as query and document identifiers.
    \item We discovered that over 20\% of the Baidu-ULTR dataset comprised only two title-abstract token combinations. To avoid these documents polluting model training, we remove the \emph{what other people searched} item from the ranking ($\approx 9\%$ of documents) and skip all documents with only a dash in the title, indicating no available content ($\approx 13\%$ of documents).\footnote{While Baidu-ULTR was tokenized with a private vocabulary, \citet{Zou2022Baidu} confirmed to us that the tokens: $[3742, 0111492, 0112169, 015061, 0116905]$ translate to \emph{what other people searched} and token $21429$ translates to ``-'' indicating missing content.} Note that documents displayed after a skipped document keep their original position and do not, e.g., move to the position of a skipped document.
    \item Third, we drop all queries with less than five documents to display, affecting $\approx 2\%$ of the train queries and $\approx 0.3\%$ of annotated queries, leaving a remainder of 6,985 annotated test queries.
\end{enumerate*}
In contrast to previous work~\citep{Niu2023-ULTRE2}, we do not remove queries without any clicks. While this is common in ULTR as sessions without clicks do not contribute to popular pairwise ranking losses~\cite{Joachims2017IPW}, they are essential to train and evaluate methods predicting calibrated click probabilities.

\vspace{-1mm}
\subsubsection{Analysis of the reranking dataset}
\label{sec:reranking}
The reranking dataset comprises three dataset partitions for training ($\approx 1.8$ million user sessions, $\approx 11.7$ million query-document pairs) and one partition for validation and testing ($\approx 590k$ user sessions, $\approx 4.8$ million query-document pairs). Table~\ref{tab:stats} gives an overview of the statistics of the reranking dataset. While the dataset is a fraction of Baidu-ULTR, we highlight that it is still substantially larger than existing \acs{ULTR} datasets~\citep{Ai2018Tiangong} and contains more unique queries and query-document pairs than the LTR datasets commonly used in semi-synthetic simulation for \acs{ULTR}~\citep{Dato2022Istella22,Dato2016Istella,Qin2013MSLR,Chapelle2011Yahoo}.

Besides query-document text and identifiers (as described in the preprocessing section), the reranking dataset also contains three sets of pre-computed query-document features: the CLS token of the BERT cross-encoder released by \citet{Zou2022Baidu}, the CLS token of a BERT cross-encoder trained by us in the same fashion as the original model, and classic lexical \acf{LTR} features, including BM25~\citep{Robertson1995BM25}, Tf-IDF~\citep{Robertson2004TfIDF}, and query-likelihood (with Jelinek-Mercer and Dirichlet smoothing~\citep{Chen1999QLSmoothing}). We compute the required inverted index for the methods above on the \emph{language modeling dataset} that we use to train our BERT cross-encoder for fair comparison. As we restrict ourselves to solely training on click data and only use the expert annotations in the final evaluation, we compute all lexical matching features with default parameters instead of tuning them on annotations. Also, LTR parameters published by ~\citet{Chen2023-THUIR-ULTR-2} do not lead to major improvements over untuned default parameters in our setting. We publish two versions of the dataset on Huggingface, one using the Baidu cross-encoder\footnote{\url{https://huggingface.co/datasets/philipphager/baidu-ultr_baidu-mlm-ctr}} and one with our cross-encoder\footnote{\url{https://huggingface.co/datasets/philipphager/baidu-ultr_uva-mlm-ctr}} with a list of all pre-computed lexical ranking features and their respective hyperparameters.

As the reranking dataset is a random sample of the larger (preprocessed) dataset, we can use it to analyze the basic properties of Baidu-ULTR. After preprocessing, the average search session contains $8$ documents, with an average of $0.68$ clicks per session (number of clicks/number of sessions) and $\approx 46.5 \%$ of sessions containing at least one click (see Table~\ref{tab:stats}). Notably, we measure the overlap between queries (md5 hash of query tokens) in the click dataset and the annotated test set and find that only $12.7\%$ of annotated queries occur in the training dataset. While they appear during training, they make up $< 0.1\%$ of the training dataset.

\begin{table}[t]
    \centering
    \caption{Description of the reranking dataset we derive from Baidu-ULTR~\citep{Zou2022Baidu}; statistics after preprocessing.}
    \begin{tabular}{l rrr}
    \toprule
         &  & \multicolumn{2}{c}{Implicit feedback} \\
        \cmidrule{3-4}
        & Annotations & train & test \\
        \midrule
        Unique queries & $6,985$ & $1,378,901$ & $501,215$  \\
        Unique documents & $381,552$ & $9,455,953$ & $3,557,825$ \\
        Total query/doc pairs & $382,038$ & $11,715,447$ & $4,209,900$ \\ 
        Total sessions & - & $1,779,017$ & $593,930$ \\
        Total impressions & - & $14,526,276$ & $4,848,878$ \\
        \midrule
        \multicolumn{2}{l}{Avg. docs per session} & \multicolumn{2}{c}{$8.165\phantom{\%}$} \\
        \multicolumn{2}{l}{Avg. clicks per session} & \multicolumn{2}{c}{$0.688\phantom{\%}$} \\
        \multicolumn{2}{l}{Avg. clicks per document} & \multicolumn{2}{c}{$0.084\phantom{\%}$} \\
        \multicolumn{2}{l}{\% of sessions with $\geqslant1$ click} & \multicolumn{2}{c}{$46.581\%$} \\
        \multicolumn{2}{l}{\% of sessions with $\geqslant2$ clicks} & \multicolumn{2}{c}{$13.082\%$} \\
        \multicolumn{2}{l}{\multirow{2}{*}{\shortstack[l]{\% of train queries occurring in the\\ annotated set}}} & \multicolumn{2}{c}{\multirow{2}{*}{$\phantom{1}0.064\%$}} \\
        \multicolumn{2}{l}{} & \multicolumn{2}{c}{} \\
        \multicolumn{2}{l}{\multirow{2}{*}{\shortstack[l]{\% of annotated queries occurring in \\ the train set}}} & \multicolumn{2}{c}{\multirow{2}{*}{$12.713\%$}} \\
        \multicolumn{2}{l}{} & \multicolumn{2}{c}{} \\
    \bottomrule
    \end{tabular}
    \label{tab:stats}
    \vspace{-5mm}
\end{table}

\vspace{-1mm}
\subsection{Position bias}
\label{sec:position_bias}
Standard counterfactual LTR methods using IPS require an estimation of position bias, which we estimate on our reranking dataset. We implement three intervention harvesting techniques that leverage the co-occurrence of the same query-document pair at different ranks~\cite{Agarwal2019AllPairs}. Intervention harvesting mines query-document pairs logged in various positions, ideally due to distinct rankers in an A/B test ranking items differently. We can use this natural variability of encountering a document in different positions to estimate bias using click ratios between neighboring positions (Adjacent Pair), each position and a fixed rank (Pivot Rank), or between all ranking positions (All Pairs). We refer the reader to \citet{Agarwal2019AllPairs} for a detailed introduction to intervention harvesting.

\begin{figure}
    \centering
    \includegraphics[width=.9\linewidth]{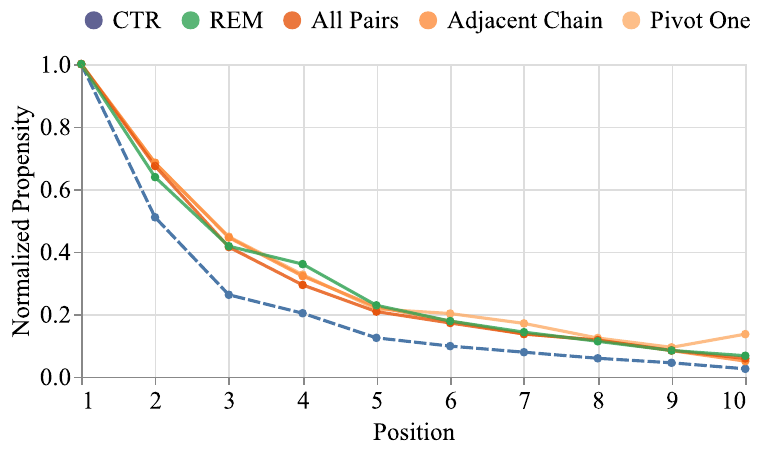}
    \caption{Position bias as estimated by RegressionEM and three intervention harvesting methods compared to the mean CTR. Propensities were normalized by position one.}
    \label{fig:position-bias}
    \vspace*{-0.5cm}
\end{figure}

We also estimate position bias using RegressionEM (REM)~\citep{Wang2018RegressionEM}, which leverages query-document embeddings. REM estimates position bias over co-occurrences of query-document pairs with similar features rather than strictly identical query-document pairs, like intervention harvesting. In our case, we use the semantic embeddings of our naive BERT cross-encoder. Figure~\ref{fig:position-bias} displays our propensity estimations and the click-through rate per rank. The estimated position bias is broadly consistent across methods. REM and intervention harvesting arrive at similar estimations from different query-document representations. Our finding suggests the presence of a noticeable, top-heavy position bias affecting user behavior on Baidu-ULTR for which \acs{ULTR} methods should be helpful.

\section{Unbiased Learning-to-Rank Methods}
\label{sec:method}

This section introduces the different ULTR methods used throughout this work. Similarly to \citet{Zou2022Baidu}, we restrict ourselves to methods that address position bias according to the \acs{PBM} and focus on benchmarking the learning algorithm used by each method.

First, we introduce the notation we will use to describe each method. Let $q$ be a user query and $D_q$ the ordered list of documents for $q$. Let $d \in D_q$ be a document in the list displayed at position $k$. We use $R_{q,d}$, $E_k$ and $C_{q,d,k}$, respectively, to denote the random events for:
\begin{enumerate*}[label=(\roman*)]
    \item document $d$ being relevant for query $q$,
    \item position $k$ of the result page being examined by the user, and 
    \item document $d$ placed in position $k$ for query $q$ being clicked.
\end{enumerate*}
We assume users follow the \acf{PBM}, meaning that users click only if they observed the position $k$ of an item and deemed the displayed document $d$ to be relevant:
\begin{equation}
    P(C_{q,d,k}) = P(E_k) \times P(R_{q,d}).
\end{equation}
In the following, we use $\mathcal{L}$ to reference a ranking loss function; in this work, either binary cross-entropy, softmax cross-entropy~\citep{Bruch2019Softmax}, or LambdaRank~\citep{Burges2006Lambdarank}. All ranking methods aim to estimate document relevance $r(q,d) = P(R_{q,d})$ and optionally examination $e(k) = P(E_k)$ (or related quantities) from observed clicks $c \in \{0, 1\}$. We use $\tilde{x}$ to denote estimators of a quantity $x$. We can now define the methods compared in our study:

\noindent\textbf{Naive}: is the common ULTR term for applying a ranking loss without any position bias correction. This means that we naively interpret clicks as positive/negative user feedback:
    \begin{equation}
        \mathcal{L}^{\textup{Naive}} = \mathcal{L}(\tilde{c}(q,d);c).
    \end{equation}
    We implement a pointwise version of this model using binary cross-entropy and two listwise versions using softmax cross-entropy~\citep{Bruch2019Softmax} and LambdaRank~\citep{Burges2006Lambdarank}, respectively.
    
\noindent\textbf{Two-Tower}~\citep[Two-Tower Model]{Yan2022TwoTowers}: jointly learns the maximum-likelihood parameters of a relevance model $\tilde{r}$ and an examination model $\tilde{e}$, directly mirroring the PBM assumptions:
    \begin{equation}
        \begin{split}
            \mathcal{L}^{\textup{TwoTower}} = \mathcal{L}\left(\sigma\left(\tilde{e}(k) + \tilde{r}(q,d)\right); c\right).
        \end{split}
    \end{equation}
    We use the additive formulation~\cite{Yan2022TwoTowers,Zhang2023DisentanglingTwoTowers} with $\tilde{r}$ and $\tilde{e}$ representing logits, $\sigma$ the sigmoid function, and $\mathcal{L}$ the binary cross-entropy loss.
    
\noindent\textbf{RegressionEM} \citep[Regression expectation-maximization]{Wang2018RegressionEM}: learns a maximum likelihood estimate of relevance and examination parameters through expectation-maximization~\citep{Dempster1977EM}:
    \begin{equation}
        \begin{split}
            \mathcal{L}^{\textup{REM}} ={}& \mathcal{L}(\tilde{r}(q,d); c + (1-c)\overline{r}(q,d) ) \\
                                       & + \mathcal{L}\left(\tilde{e}(k); c+ (1-c)\overline{e}(k)\right),
        \end{split}
    \end{equation}
    where $\overline{r}(q,d) = \frac{\tilde{r}(q,d)\left( 1 - \tilde{e}(k)\right)}{1 - \tilde{r}(q,d) \tilde{e}(k)}$ and $\overline{e}(k) = \frac{\tilde{e}(k)\left( 1 - \tilde{r}(q,d)\right)}{1 - \tilde{r}(q,d) \tilde{e}(k)}$ are the posterior relevance and examination probabilities. In practice, we use binary cross-entropy as the base loss and replace the original expectation-maximization procedure with a numerically stable gradient-based version suggested in TF-Rank~\citep{Kumar2019TFRank} by performing the posterior computation with logits.
    
\noindent\textbf{IPS} \citep[Inverse propensity scoring]{Joachims2017IPW}: Given known examination probabilities $e_k$, IPS re-weights the click labels by the propensity (i.e., normalized examination probability) of the current position:
    \begin{equation}
        \mathcal{L}^{\textup{IPS}} = \mathcal{L}\left(\tilde{r}(q,d); \frac{\text{max}(\tau, e_1)}{\text{max}(\tau, e_k)} c\right).
    \end{equation}
    In our experiments, we use propensities estimated by the AllPairs intervention harvesting method~\citep{Agarwal2019AllPairs} and reduce variance by clipping examination probabilities to a minimum value of $\tau = 0.1$~\citep{Joachims2017IPW}. We implement a pointwise~\citep{Bekker2019PointwiseIPS,Saito2020PointwiseIPS} and listwise IPS variant~\cite{Bruch2019Softmax,Ai2018DLA}.
    
\noindent\textbf{DLA} \citep[Dual learning algorithm]{Ai2018DLA}: separately learns tunable relevance scores with fixed propensities and tunable propensities with fixed relevance scores, essentially applying IPS twice:
    \begin{equation}
        \mathcal{L}^{\textup{DLA}} = \mathcal{L}\left(\tilde{r}(q,d); \frac{\tilde{e}(1)}{\tilde{e}(k)} c \right) + \mathcal{L}\left(\tilde{e}(k); \frac{\tilde{r}(q,d_1)}{\tilde{r}(q,d)} c\right),
    \end{equation}
    where $d_1$ is the document ranked first on the current result page. We use softmax cross-entropy as the base loss and perform a softmax-normalization of examination and relevance probabilities $\tilde{r}$ and $\tilde{e}$ within a result page, as in the original paper~\citep{Ai2018DLA}.
    
\noindent\textbf{PairD} \citep[Pairwise debiasing]{Hu2019UnbiasedLambdaMART}: learns positive $\tilde{e}^+$ and negative $\tilde{e}^-$ propensities (i.e., corresponding to clicks and non-clicks, respectively) with a constraint on the norm of the learned propensities:
    \begin{equation}
        \mathcal{L}^{\textup{PairD}} = \frac{\mathcal{L}(\tilde{r}(q,d); c)}{\tilde{e}^+(k) \tilde{e}^-(k)} + \| \tilde{e}^+ \|_1 + \| \tilde{e}^- \|_1.
    \end{equation}
    In practice, we use the $L_1$-norm for propensity regularization and learn the relevance scores using the LambdaRank loss~\citep{Burges2006Lambdarank}, as in the original paper~\citep{Hu2019UnbiasedLambdaMART}. Note that the theoretical validity of this method under non-trivial position bias has been challenged by~\citet{Oosterhuis2022Limitations}. Still, we include it in our comparison as it has demonstrated strong empirical performance in past comparisons~\citep{Ai2021ULTROnlineOffline}.

\section{Experimental Setup}
\label{sec:setup}
\subsection{Training and evaluation procedure}

All models are trained on our reranking dataset, and we use a 50/50 random split of our test click dataset (see Table~\ref{tab:stats}) for validation and testing. We evaluate ranking performance on $6,985$ annotated queries, where experts rated each query-document pair's relevance on a scale from $0$ to $4$. We point to \citet{Zou2022Baidu} for more details on the annotation process. In this work, we do not use available side information such as dwell-time, off-screen scrolling, and returns to the result page. Instead, we focus on query and document content, document position and click label as our training data.

We measure ranking performance on the annotated set using \acf{DCG} at different truncation levels and \acf{MRR} at $10$, as well as \acf{NLL} for click prediction.

\vspace{-2mm}
\subsection{Models}
Traditionally, \acl{ULTR} practitioners train small ranking models with LTR features such as BM25~\citep{Robertson1995BM25} or TF-IDF~\citep{Robertson2004TfIDF} as input~\citep{Ai2018DLA,Oosterhuis2021LambdaLoss,Vardasbi2020Affine}. To investigate the role of semantic embeddings and the interaction of \acs{ULTR} with language model training on such a large dataset, we train two types of models: transformer-based \emph{language models} trained from scratch following the MonoBERT cross-encoder~\citep{Nogueira2019MonoBERT}, and multi-layer perceptron-based \emph{reranking models}, that take as input either traditional \acs{LTR} features or fixed query-document embeddings obtained by the language models. 

\subsubsection{Language models}
We train cross-encoders on the Baidu-ULTR dataset from scratch in Jax~\citep{Bradbury2018Jax}, building on the FlaxBERT implementation from Huggingface.\footnote{\url{https://huggingface.co/docs/transformers/model_doc/bert\#transformers.FlaxBertForPreTraining}} The input to our BERT models is: \texttt{[CLS] query [SEP] doc\_title [SEP] doc\_abstract}. Following~\citep{Zou2022Baidu,Chen2023-TENCENT-ULTR-1}, we truncate the input to a maximum length of 128 tokens. Compared to the original BERT training scheme~\citep{Devlin2019BERT}, we keep the masked-language modeling task (masking $30\%$ of input tokens at random) but discard the next-sentence prediction task. Instead, we follow the MonoBERT~\citep{Nogueira2019MonoBERT} setup and apply a linear layer to the BERT CLS token to output a click prediction score. We train multiple models with different click-based loss functions: naive click prediction with a pointwise (binary cross-entropy) and a listwise (softmax cross-entropy) loss, pointwise and listwise \acs{IPS}, a pointwise two-tower loss, and a listwise \acs{DLA} loss (see Section~\ref{sec:method}). All loss functions were built on top of the Rax library~\citep{Jagerman2022Rax}.

We keep the architecture of the original $\text{BERT}_{\text{base}}$ model, using 12 transformer layers, 12 attention heads, 768 output dimensions, and, following \citet{Chen2023-TENCENT-ULTR-1}, a vocabulary size of $22,000$. All language models are trained with a batch size of $256$ ($4 \times 64$) for $2$ million gradient steps, i.e., on $512$ million documents. Training each base model on four GPUs (NVIDIA H100-80GB) takes around 46 hours, which is a speed-up of almost 50\% compared to an early PyTorch implementation of ours using the exact same library, architecture, and hardware. We use the AdamW optimizer~\citep{Loshchilov2019AdamW} with a fixed learning rate of $5 \times 10^{-5}$ and a weight decay of $0.01$. Given the substantial compute invested to train each language model, we rely on prevalent default parameters for BERT (listed in our repository).

\subsubsection{Reranking models}
To investigate the interactions of \acs{ULTR} with semantic query-document embeddings and traditional \acs{LTR} features on Baidu-ULTR, we train several smaller feed-forward networks as reranking models, as is common in the \acs{ULTR} community~\citep{Ai2018DLA,Vardasbi2020Cascade,Oosterhuis2021LambdaLoss}. Our reranking models are composed of linear layers with ReLU activations and, optionally, dropout regularization. We found no benefit in applying Layer or Batch Normalization, as our BERT embeddings are already normalized. As \acs{LTR} features can span a large range, we find that scaling features with $\text{log1p}(x) = \log_e(1+|x|) \odot \operatorname{sign}(x)$ before the first layer, as suggested by \citet{Qin2021AreNeuralRankers}, works well. Note that the feed-forward network takes the query-document features as input and, depending on the ULTR method, outputs a relevance or click prediction. We add a single learnable model parameter per position for methods that jointly estimate position bias. In contrast to models leveraging multiple bias features~\cite{Zhao2022-ULTRE1,Yan2022TwoTowers}, we found no additional benefit by using a multi-layer perceptron for position bias estimation in our setting.

\vspace{-3mm}
\subsection{Hyperparameter tuning}
We perform extensive hyperparameter tuning for a fair comparison of the reranking models in our experiments. Given the immense combinatorial space of hyperparameters, methods, and datasets, we adopt an incremental hyperparameter tuning strategy as advocated by~\citet{Godbole2023TuningPlaybook}. First, we tune our model architecture per set of input features based on the pointwise naive model. We tune the number of hidden dimensions $\in \{64, 128, 256, 512, 1024\}$ and the number of layers $\in \{2, 3, 4, 5\}$ over three random seeds. As both network depth and width can interact with the learning rate, we tune each parameter combination over three learning rates $\in \{0.001, 0.0005, 0.0001\}$, i.e., we treat the learning rate as a nuisance parameter~\citep{Godbole2023TuningPlaybook}. We adopt this incremental tuning procedure as we found no major discrepancies in model architecture between ULTR methods but instead between sets of input features. Subsequently, we adopt the architecture of the pointwise naive model per dataset for all other methods.

A five-layer perceptron with 512 hidden dimensions yields optimal results for our LTR features, while a five-layer perceptron with 256 dimensions performs best for the Baidu BERT embeddings. Additionally, a two-layer perceptron with 256 dimensions was the most suitable choice for our BERT embeddings. Given each base model architecture, we tune the final dropout $\in \{0, 0.3\}$ and learning rate $\in \{0.001, 0.0005, 0.0003, 0.0001\}$ for each method and dataset over three random seeds. We use the AdamW optimizer~\citep{Loshchilov2019AdamW} with $\beta_1 = 0.9$, $\beta_2 = 0.999$, and $\varepsilon = 1e^{-8}$. Most models work best with a learning rate of $0.0001$ and do not benefit from dropout. The full list of hyperparameters is available in our online repository. Given the final hyperparameters for each method, we train all methods for 50 epochs, stopping early after five epochs of no improvement of the validation loss computed on clicks. Lastly, while this tuning step improved the performance of all methods, our main findings are consistent across many different hyperparameter combinations.

\section{Results}
\label{sec:results}

\begin{figure*}[t]
    \centering
    \includegraphics[width=\linewidth]{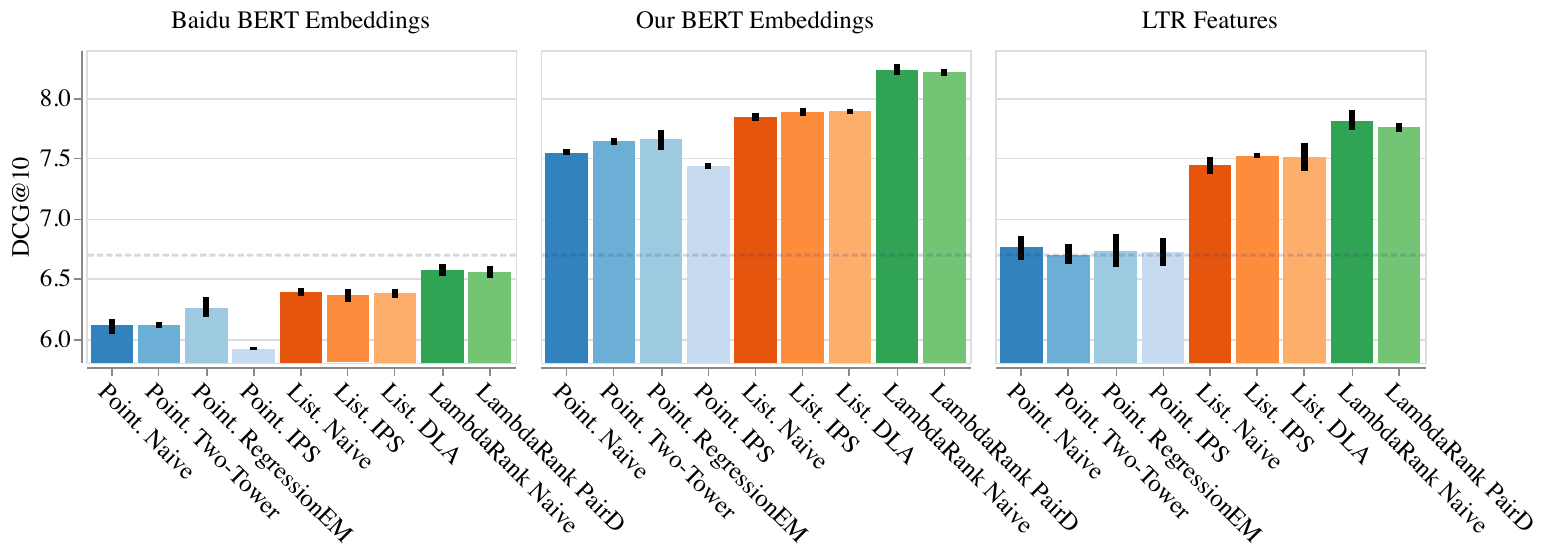}
    \caption{Comparing ULTR methods on pre-trained BERT embeddings and LTR features. We display the average ranking performance measured in DCG@10 over five independent runs and plot a bootstrapped 95\% confidence interval. The grey dotted line indicates the performance of a random ranker.}
    \label{fig:ultr-on-our-bert}
    \vspace{-4mm}
\end{figure*}

Before presenting our findings, we recall our research questions:
\begin{enumerate}[nosep, leftmargin=*, label = \textbf{(RQ\arabic*)}]
    \item Does \acl{ULTR} improve performance on the Baidu-ULTR dataset over naive, non-debiasing models?
    \item How do \acs{ULTR} methods fare against each other, and how do ranking losses and input features affect their performance?
    \item Can \acs{ULTR} methods be applied during language model training and do they bring improved performance?
\end{enumerate}

\vspace{-3mm}
\subsection{Unbiased \acl{LTR} yields no or tiny improvements on Baidu-ULTR}
\label{sec:ultr-does-not-work}

In our first experiment, we train reranking models with pointwise and listwise \acs{ULTR} loss functions and their respective naive, non-debiasing counterparts on three different types of input features: the CLS token of~\citet{Zou2022Baidu}'s pointwise cross-encoder, the CLS token of our pointwise cross-encoder, and \acl{LTR} features (see Section~\ref{sec:reranking}). We list comprehensive results in Table~\ref{tab:results}. First, we focus on the ranking performance on expert annotations as measured in \acs{DCG} and \acs{MRR}. We display the \acs{DCG}@10 in Figure~\ref{fig:ultr-on-our-bert}.

At first glance at Figure~\ref{fig:ultr-on-our-bert}, echoing previous work~\citep{Zhao2022-ULTRE1,Niu2023-ULTRE2}, we observe that \emph{we can get a much higher ranking performance from our cross-encoder than the initially released BERT cross-encoder}. Models trained on \acs{LTR} features and even of a trivial baseline assigning random scores (grey dotted line) to the annotated documents beat all models trained on the original Baidu BERT features.

Second, given a set of input features, the choice of ranking loss used as a base for \acs{ULTR} methods matters greatly (comparing the colored groups in Figure~\ref{fig:ultr-on-our-bert}). Methods based on LambdaRank outperform methods based on the listwise softmax loss, which outperform those based on the pointwise binary cross-entropy loss. In contrast, \emph{applying \acs{ULTR} techniques yields marginal ranking improvements at most compared to their naive, non-debiasing counterpart}.

In detail, we observe that the pointwise two-tower and \acs{IPS} models do not consistently and significantly improve performance compared to the naive pointwise loss -- \acs{IPS} is even significantly worse on both BERT features -- and RegressionEM is the only pointwise \acs{ULTR} method that brings small but sometimes significant improvements. Regarding listwise methods, \acs{IPS} and \acs{DLA} significantly but marginally improve on our BERT and LTR features. Finally, PairD is no better and sometimes worse than its naive LambdaRank counterpart. \emph{These results are broadly consistent across input features, suggesting that standard \acs{ULTR} methods struggle to bring improvement to Baidu-ULTR, regardless of the query-document features.}

Overall, we can answer research questions (RQ1) and (RQ2): on the Baidu-ULTR dataset, \acl{ULTR} methods do not consistently improve ranking performance on expert annotations, particularly when contrasted with the significant and reliable differences based on the choice of query-document features and ranking loss. Our reranking results confirm the findings of \citet{Zou2022Baidu}.

\vspace{-3mm}
\subsection{Language model training is sensitive to the choice of \acl{ULTR} method}
\label{sec:language-model-results}

Given the poor performance of \acs{ULTR} on our reranking datasets (see Section~\ref{sec:ultr-does-not-work}), we further investigate whether training language models with \acs{ULTR} is a promising direction for future research. We train six MonoBERT~\citep{Nogueira2019MonoBERT} cross-encoders with different pointwise and listwise loss functions, including \acs{ULTR} loss functions.

Table~\ref{tab:base-models} gives an overview of applying \acs{ULTR} methods during language model training. In contrast to the inefficacy of \acs{ULTR}-based rerankers, \emph{we observe stark differences when applying \acs{ULTR} methods during language model training}. Similar to the reranking task, approaches based on the listwise softmax loss perform better than those based on pointwise binary cross-entropy. The additional application of \acs{ULTR} brings considerable improvements with the pointwise two-tower objective. However, we also observe substantial degradations with both IPS and DLA. The best method is the naive listwise softmax loss without any debiasing objective.

These inconclusive results make us cautious about answering (RQ3): \acs{ULTR} methods substantially impact language model training more than reranking with fixed embeddings. However, the influence of \acs{ULTR} on ranking performance and learned query-document representations needs further investigation.

\vspace{-3mm}
\subsection{Click prediction does not imply ranking performance on annotations}

\begin{figure}
    \centering
    \includegraphics[width=.9\linewidth]{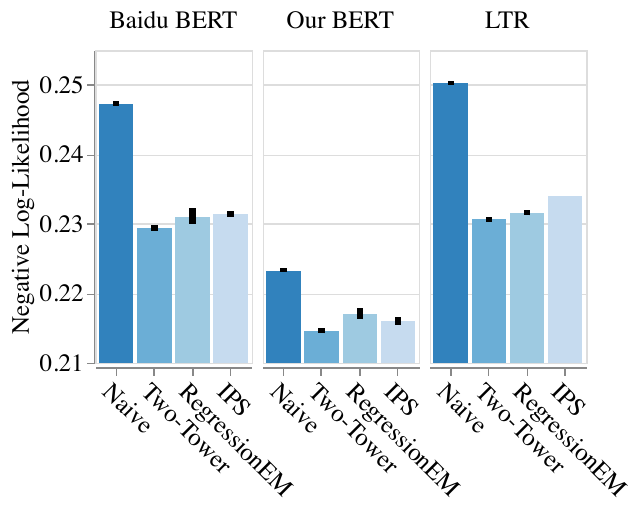}
    \caption{Click prediction performance of pointwise methods measured in negative log-likelihood; lower is better.}
    \label{fig:nll-vs-dcg}
    \vspace{-5mm}
\end{figure}

In addition to evaluating ranking performance, we display the \acf{NLL} of predicted click scores in Figure~\ref{fig:nll-vs-dcg}. We restrict this evaluation to pointwise ULTR reranking methods, making well-defined click predictions. All \acs{ULTR} methods robustly improve click prediction compared to a naive loss, with the two-tower model showing the largest improvement on all three datasets. Moreover, as discussed in Section~\ref{sec:position_bias}, the propensities learned by RegressionEM are close to those discovered through intervention harvesting and show a strong position bias. This finding suggests that \acl{ULTR} robustly captures position bias and can better predict user clicks because of it. \emph{Yet, this improved click prediction does not translate to enhanced ranking performance on expert annotations.}

Critically, we can see in Table~\ref{tab:results} that ranking documents according to their  (untuned) BM25~\citep{Robertson1995BM25} scores yields better ranking performance than all click-based models we trained. Even more surprisingly, using BM25 scores as part of the input for the reranking models -- it is one of the \acs{LTR} features -- and training these models on clicks lowers ranking performance. This phenomenon was also described by \citet{Sun2023-CIR-PRETRAIN-3} during the WSDM Cup.
These concurrent results suggest that \emph{click prediction and ranking performance on annotations are diverging objectives on Baidu-ULTR and training models on clicks does not guarantee improvements in ranking metrics.}

\section{Discussion}

In light of the puzzling results reported in Section~\ref{sec:results}, we review potential reasons for the failure of~\acs{ULTR} in Section~\ref{sec:potential-explanations} as well as the limitations of our study in Section~\ref{sec:limitations}. Finally, we discuss the implications for the field of \acl{ULTR} in Section~\ref{sec:implications}.

\begin{table}[t]
    \caption{Comparison of cross-encoder models trained from scratch on the Baidu-ULTR dataset with \acs{ULTR} loss functions.}
    \label{tab:base-models}
    \centering
    \begin{tabular}{l cc}
        \toprule
        Model & DCG@10 $\uparrow$ & NLL $\downarrow$ \\
        \midrule
        Pointwise Naive & $7.251$ & $0.227$  \\
        Pointwise Two-Tower & $7.456$ & $0.218$ \\
        Pointwise IPS & $6.296$ & $0.222$ \\
        \midrule
        Listwise Naive & $8.478$ & - \\
        Listwise IPS & $7.450$ & - \\
        Listwise DLA & $7.802$ & - \\
        \bottomrule
    \end{tabular}
    \vspace{-4mm}
\end{table}

\subsection{Potential reasons for the failure of \acs{ULTR}}
\label{sec:potential-explanations}

\textbf{No position bias.} A first straightforward explanation for the inefficacy of \acs{ULTR} techniques would be that position bias is not prevalent in the Baidu-ULTR dataset. However, our results in Section~\ref{sec:position_bias} show a strongly decreasing \acl{CTR} with increasing position and a strong position bias estimated by techniques based on different methodologies. Recall that intervention harvesting uses \acs{CTR} ratios of query-document pairs appearing in multiple positions while RegressionEM uses our BERT features and expectation maximization. The fact that these results agree strongly suggests the existence of position bias in this dataset and that \acs{ULTR} methods like RegressionEM and Two-Tower models could capture it.

\noindent\textbf{More complex user behavior.} \citet{Zou2022Baidu} suggest that the actual user model might be more complicated than the \acs{PBM} considered in this work. This hypothesis might explain why listwise approaches bring such considerable improvements over pointwise methods and would not be surprising as user studies have identified more complex biases~\citep{Agarwal2019TrustBias,Zhuang2021XPA,Sarvi2023Outliers}. However, \acs{ULTR} methods bringing no benefits would mean that the substantial position bias we identified is negligible against other types of biases. This assumption, in turn, would contradict related work where \acs{PBM}-based models trained on more complex user behaviors still improved performance, albeit by not as much as the correct user model~\citep{Vardasbi2020Cascade, Deffayet2023Robustness}.

\noindent\textbf{Lack of identifiability.} Past studies have identified that a lack of variability in the logged data, i.e., not encountering a query-document pair at different positions, can lead to the dataset not being identifiable, in the sense that there exist infinite valid combinations of relevance and bias parameters~\citep{Oosterhuis2022Limitations, Chen2024Identifiability}. However, as this work's four bias estimation methods converged to a similar position bias, we deem this explanation rather unlikely.

\noindent\textbf{Distribution shift.} Poor ranking performance may also be caused by the distribution shift between the query-document pairs in the training and annotated test set. While methods train only on the top-$10$ search results, the annotation dataset also includes presumably much less relevant documents sampled from the top-$1000$ candidate documents. This shift in the input feature distribution, in conjunction with the low overlap between train and test queries (see Section~\ref{sec:reranking}) and the long-tail query distribution during training, might all potentially impact \acs{ULTR}. More research and, ideally, clicks for queries with expert annotations are needed to better understand the impact of the various distribution shifts on \acs{ULTR}.

\noindent\textbf{Strong logging policy.} While we do not have access to logging policy scores and rankings on the annotated test queries, the limitations imposed by an already strong production system have been well-documented~\citep{Deffayet2023CMIP}. A side observation of ours reinforces this possibility: methods that show strong ranking performance correlate highly with the logging policy of the click dataset (as estimated through the original item position). For future dataset releases, we recommend including logging policy scores to enable assessments of the improvement made by training new rankers on click data compared to the logging policy~\citep{Deffayet2023CMIP,Gupta2023SafeDeployment}.

\noindent\textbf{User-annotator disagreement.} Finally, the divergence of click prediction and ranking performance hints at a deeper issue: annotators may not find the same documents relevant as users scrolling through a result page, which might be personalized and part of a multi-query session to find actionable information for the current user needs. We hope further analysis of Baidu-ULTR or parallel datasets of clicks and expert annotations can clarify this hypothesis.

\begin{table*}[th]
\centering
\vspace*{-2mm}
\caption{
Comparison of ULTR methods on pre-trained BERT embeddings and LTR features, displaying averaged results over five independent runs with standard deviation in parentheses. $\uparrow$ indicates the higher the better and $\downarrow$ the lower the better. Methods are grouped by loss, and significant differences are marked with $\blacktriangle$ or $\blacktriangledown$ compared to the naive method in each group using a two-sided paired t-test with $\alpha = 0.01$ and Bonferroni correction.}
\label{tab:results}
\begin{tabular}{llllllll}
    \toprule
    Input features & Method & DCG@1 $\uparrow$ & DCG@3 $\uparrow$ & DCG@5 $\uparrow$ & DCG@10 $\uparrow$ & MRR@10 $\uparrow$ & NLL $\downarrow$ \\
    \midrule 
     & Random & $1.472$ \scriptsize{$(0.020)$} & $3.148$ \scriptsize{$(0.030)$} & $4.349$ \scriptsize{$(0.036)$} & $6.693$ \scriptsize{$(0.044)$} & $0.577$ \scriptsize{$(0.003)$} & $0.944$ \scriptsize{$(0.002)$} \\
     & BM25 $(k_1 = 1.2, b = 0.75)$ & $2.211$ & $4.656$ & $6.377$ & $9.544$ & $0.716$ & - \\
    \midrule
    \multirow{9}{*}{Baidu BERT} & Pointwise Naive & $1.264$ \scriptsize{$(0.030)$} & $2.765$ \scriptsize{$(0.059)$} & $3.881$ \scriptsize{$(0.068)$} & $6.111$ \scriptsize{$(0.079)$} & $0.535$ \scriptsize{$(0.007)$} & $0.246$ \scriptsize{$(0.005)$} \\
     & Pointwise Two-Tower & $1.266$ \scriptsize{$(0.018)$} & $2.769$ \scriptsize{$(0.023)$} & $3.893$ \scriptsize{$(0.026)$} & $6.114$ \scriptsize{$(0.031)$} & $0.533$ \scriptsize{$(0.003)$} & $0.228$ \scriptsize{$(0.005)$}$^\blacktriangledown$ \\
     & Pointwise RegressionEM & $1.343$ \scriptsize{$(0.034)$}$^\blacktriangle$ & $2.884$ \scriptsize{$(0.074)$}$^\blacktriangle$ & $4.017$ \scriptsize{$(0.091)$}$^\blacktriangle$ & $6.257$ \scriptsize{$(0.107)$}$^\blacktriangle$ & $0.548$ \scriptsize{$(0.005)$}$^\blacktriangle$ & $0.229$ \scriptsize{$(0.005)$}$^\blacktriangledown$ \\
     & Pointwise IPS & $1.157$ \scriptsize{$(0.011)$}$^\blacktriangledown$ & $2.606$ \scriptsize{$(0.012)$}$^\blacktriangledown$ & $3.699$ \scriptsize{$(0.013)$}$^\blacktriangledown$ & $5.915$ \scriptsize{$(0.014)$}$^\blacktriangledown$ & $0.517$ \scriptsize{$(0.002)$}$^\blacktriangledown$ & $0.230$ \scriptsize{$(0.005)$}$^\blacktriangledown$ \\
     \cmidrule{2-8}
     & Listwise Naive & $1.362$ \scriptsize{$(0.022)$} & $2.940$ \scriptsize{$(0.037)$} & $4.096$ \scriptsize{$(0.037)$} & $6.388$ \scriptsize{$(0.042)$} & $0.549$ \scriptsize{$(0.004)$} & - \\
     & Listwise IPS & $1.353$ \scriptsize{$(0.034)$} & $2.920$ \scriptsize{$(0.060)$} & $4.078$ \scriptsize{$(0.063)$} & $6.359$ \scriptsize{$(0.071)$}$^\blacktriangledown$ & $0.548$ \scriptsize{$(0.005)$} & - \\
     & Listwise DLA & $1.345$ \scriptsize{$(0.020)$} & $2.917$ \scriptsize{$(0.037)$} & $4.083$ \scriptsize{$(0.048)$} & $6.378$ \scriptsize{$(0.051)$} & $0.549$ \scriptsize{$(0.004)$} & - \\
     \cmidrule{2-8}
     & LambdaRank Naive  & $1.419$ \scriptsize{$(0.017)$} & $3.047$ \scriptsize{$(0.039)$} & $4.237$ \scriptsize{$(0.054)$} & $6.570$ \scriptsize{$(0.065)$} & $0.562$ \scriptsize{$(0.005)$} & - \\
     & LambdaRank PairD & $1.410$ \scriptsize{$(0.028)$} & $3.025$ \scriptsize{$(0.044)$} & $4.215$ \scriptsize{$(0.064)$} & $6.550$ \scriptsize{$(0.066)$} & $0.559$ \scriptsize{$(0.005)$} & - \\
    \midrule
    \multirow{9}{*}{Our BERT} & Pointwise Naive & $1.705$ \scriptsize{$(0.013)$} & $3.602$ \scriptsize{$(0.022)$} & $4.944$ \scriptsize{$(0.023)$} & $7.546$ \scriptsize{$(0.031)$} & $0.619$ \scriptsize{$(0.002)$} & $0.221$ \scriptsize{$(0.005)$} \\
     & Pointwise Two-Tower & $1.656$ \scriptsize{$(0.017)$}$^\blacktriangledown$ & $3.556$ \scriptsize{$(0.024)$}$^\blacktriangledown$ & $4.948$ \scriptsize{$(0.032)$} & $7.639$ \scriptsize{$(0.038)$}$^\blacktriangle$ & $0.615$ \scriptsize{$(0.004)$}$^\blacktriangledown$ & $0.213$ \scriptsize{$(0.006)$}$^\blacktriangledown$ \\
     & Pointwise RegressionEM & $1.694$ \scriptsize{$(0.031)$} & $3.619$ \scriptsize{$(0.056)$} & $4.998$ \scriptsize{$(0.081)$}$^\blacktriangle$ & $7.657$ \scriptsize{$(0.111)$}$^\blacktriangle$ & $0.618$ \scriptsize{$(0.004)$} & $0.215$ \scriptsize{$(0.005)$} \\
     & Pointwise IPS & $1.589$ \scriptsize{$(0.012)$}$^\blacktriangledown$ & $3.438$ \scriptsize{$(0.012)$}$^\blacktriangledown$ & $4.794$ \scriptsize{$(0.013)$}$^\blacktriangledown$ & $7.436$ \scriptsize{$(0.035)$}$^\blacktriangledown$ & $0.604$ \scriptsize{$(0.001)$}$^\blacktriangledown$ & $0.214$ \scriptsize{$(0.005)$} \\
     \cmidrule{2-8}
     & Listwise Naive & $1.798$ \scriptsize{$(0.008)$} & $3.768$ \scriptsize{$(0.020)$} & $5.167$ \scriptsize{$(0.029)$} & $7.844$ \scriptsize{$(0.046)$} & $0.631$ \scriptsize{$(0.001)$} & - \\
     & Listwise IPS & $1.816$ \scriptsize{$(0.017)$} & $3.800$ \scriptsize{$(0.027)$}$^\blacktriangle$ & $5.200$ \scriptsize{$(0.031)$}$^\blacktriangle$ & $7.885$ \scriptsize{$(0.043)$}$^\blacktriangle$ & $0.634$ \scriptsize{$(0.002)$}$^\blacktriangle$ & - \\
     & Listwise DLA & $1.816$ \scriptsize{$(0.011)$} & $3.806$ \scriptsize{$(0.022)$}$^\blacktriangle$ & $5.210$ \scriptsize{$(0.024)$}$^\blacktriangle$ & $7.890$ \scriptsize{$(0.031)$}$^\blacktriangle$ & $0.634$ \scriptsize{$(0.001)$}$^\blacktriangle$ & - \\
     \cmidrule{2-8}
     & LambdaRank Naive & $1.931$ \scriptsize{$(0.016)$} & $3.993$ \scriptsize{$(0.029)$} & $5.452$ \scriptsize{$(0.040)$} & $8.233$ \scriptsize{$(0.055)$} & $0.648$ \scriptsize{$(0.003)$} & - \\
     & LambdaRank PairD & $1.932$ \scriptsize{$(0.007)$} & $3.985$ \scriptsize{$(0.019)$} & $5.438$ \scriptsize{$(0.022)$} & $8.216$ \scriptsize{$(0.038)$} & $0.646$ \scriptsize{$(0.003)$} & - \\
    \midrule
    \multirow{9}{*}{LTR Features} & Pointwise Naive & $1.312$ \scriptsize{$(0.034)$} & $2.977$ \scriptsize{$(0.066)$} & $4.246$ \scriptsize{$(0.090)$} & $6.757$ \scriptsize{$(0.133)$} & $0.543$ \scriptsize{$(0.008)$} & $0.247$ \scriptsize{$(0.005)$} \\
     & Pointwise Two-Tower & $1.333$ \scriptsize{$(0.046)$} & $2.986$ \scriptsize{$(0.085)$} & $4.241$ \scriptsize{$(0.094)$} & $6.698$ \scriptsize{$(0.110)$}$^\blacktriangledown$ & $0.553$ \scriptsize{$(0.008)$}$^\blacktriangle$ & $0.228$ \scriptsize{$(0.006)$}$^\blacktriangledown$ \\
     & Pointwise RegressionEM & $1.397$ \scriptsize{$(0.068)$}$^\blacktriangle$ & $3.062$ \scriptsize{$(0.133)$}$^\blacktriangle$ & $4.300$ \scriptsize{$(0.166)$}$^\blacktriangle$ & $6.729$ \scriptsize{$(0.181)$} & $0.559$ \scriptsize{$(0.014)$}$^\blacktriangle$ & $0.229$ \scriptsize{$(0.005)$}$^\blacktriangledown$ \\
     & Pointwise IPS & $1.352$ \scriptsize{$(0.044)$}$^\blacktriangle$ & $3.013$ \scriptsize{$(0.095)$} & $4.258$ \scriptsize{$(0.126)$} & $6.717$ \scriptsize{$(0.154)$} & $0.560$ \scriptsize{$(0.011)$}$^\blacktriangle$ & $0.232$ \scriptsize{$(0.005)$}$^\blacktriangledown$ \\
     \cmidrule{2-8}
     & Listwise Naive & $1.599$ \scriptsize{$(0.023)$} & $3.485$ \scriptsize{$(0.042)$} & $4.845$ \scriptsize{$(0.072)$} & $7.443$ \scriptsize{$(0.100)$} & $0.596$ \scriptsize{$(0.005)$} & - \\
     & Listwise IPS & $1.641$ \scriptsize{$(0.020)$}$^\blacktriangle$ & $3.545$ \scriptsize{$(0.012)$}$^\blacktriangle$ & $4.920$ \scriptsize{$(0.020)$}$^\blacktriangle$ & $7.522$ \scriptsize{$(0.026)$}$^\blacktriangle$ & $0.602$ \scriptsize{$(0.002)$}$^\blacktriangle$ & - \\
     & Listwise DLA & $1.621$ \scriptsize{$(0.026)$} & $3.525$ \scriptsize{$(0.074)$}$^\blacktriangle$ & $4.891$ \scriptsize{$(0.101)$}$^\blacktriangle$ & $7.512$ \scriptsize{$(0.160)$}$^\blacktriangle$ & $0.599$ \scriptsize{$(0.006)$}$^\blacktriangle$ & - \\
     \cmidrule{2-8}
     & LambdaRank Naive & $1.750$ \scriptsize{$(0.035)$} & $3.717$ \scriptsize{$(0.068)$} & $5.125$ \scriptsize{$(0.083)$} & $7.810$ \scriptsize{$(0.111)$} & $0.613$ \scriptsize{$(0.007)$} & - \\
     & LambdaRank PairD & $1.723$ \scriptsize{$(0.018)$}$^\blacktriangledown$ & $3.683$ \scriptsize{$(0.034)$}$^\blacktriangledown$ & $5.089$ \scriptsize{$(0.042)$}$^\blacktriangledown$ & $7.761$ \scriptsize{$(0.046)$}$^\blacktriangledown$ & $0.608$ \scriptsize{$(0.004)$}$^\blacktriangledown$ & - \\
    \bottomrule
\end{tabular}

\end{table*}

\vspace{-3mm}
\subsection{Limitations}
\label{sec:limitations}

First, we only considered the correction of position bias under the \acl{PBM}, while other biases might have to be mitigated~\citep{Agarwal2019TrustBias,Vardasbi2020Cascade}. Second, we did not include bias features available in Baidu-ULTR beyond positions. We highlight that~\citet{Chen2023-TENCENT-ULTR-1} report improved ranking performance by considering additional bias features, including item height or media type.
Third, the variability in item position that we used for position bias estimation is likely not due to natural variability but due to unobserved confounding and algorithmic choices based on additional context information rather than a stochastic logging policy. Fourth, our work only applied \acs{ULTR} during language model pre-training and on fixed BERT embeddings for reranking. We did not yet explore multi-stage pre-training schemes, such as pre-training BERT on a naive click prediction task and later fine-tuning BERT using~\acs{ULTR}.

Lastly, we stress that our findings are strictly limited to the Baidu-ULTR dataset. While the dataset was collected from one of the most used search engines globally, it cannot be considered representative of all real-world search scenarios. Yet, we believe that results on this large-scale dataset are important for the research community.

\subsection{Implications for the field}
\label{sec:implications}

Our results starkly contrast with many studies in semi-synthetic simulation~\citep{Joachims2017IPW,Ai2021ULTROnlineOffline,Tran2021ULTRA,Saito2020PointwiseIPS,Ai2018DLA,Hu2019UnbiasedLambdaMART} and call for adjusting semi-synthetic experimental setups to reflect real-world challenges better. The richness of this dataset allows for seeding simulations with more plausible data and, therefore, might bridge the gap between simulations and reality. Moreover, the prevalence of transformer-based models and text embeddings in current IR research encourages exploring the interaction of \acs{ULTR} methods with such models. As shown in Section~\ref{sec:language-model-results}, the same techniques applied during language model training, instead of on a small re-ranking model, can yield vastly different results.

More broadly, the puzzling results we obtained, especially the apparent divergence between clicks and relevance annotations, prompt us to rethink how we measure success in~\acs{ULTR}. Expert annotations are static and might not reflect user context, so results obtained on annotated datasets can be polluted by distribution shifts or logging policy performance. In particular, we believe that, whenever possible, we should evaluate \acs{ULTR} methods on the tasks they are trained to accomplish, including relevance estimation, bias estimation, CTR maximization, fairness of exposure, etc.

Finally, we would like to stress that this paper is not a judgment on the \acs{ULTR} field. As mentioned above, results may differ on other datasets, and most \acs{ULTR} methods are theoretically justified, meaning only their validity in this scenario has been challenged. In fact, our experiments validate some intuitions formulated in the \acs{ULTR} community. For instance, the stark differences between loss functions justify the motivation of \citet{Joachims2017IPW} to connect simple user models with more powerful ranking loss functions.

\section{Conclusion}

In this work, we carefully revisited and extended the experiments conducted by~\citet{Zou2022Baidu} on the recently released Baidu-ULTR dataset. As the largest publicly available dataset comprising both click logs from a major search engine and expert annotations, this dataset constitutes a rare opportunity to assess the progress of click-based \acl{LTR}, and especially \acl{ULTR}.

Our main conclusion, however, is that while we have observed substantial improvements relating to the choice of query-document representations and ranking loss (e.g., pointwise or listwise), conventional \acl{ULTR} techniques do not bring clear improvements in ranking performance on the Baidu-ULTR dataset. Our findings confirm the work by \citet{Zou2022Baidu}, even though we used different dataset features and preprocessing, model implementations, and performed extensive hyperparameter tuning. 
Further, we observed a divergence between click-based and annotations-based objectives, as all click-based approaches were outperformed on annotation-based metrics by simple baselines such as BM25, even when BM25 was a model input. We believe these results call for more research into the conditions for success and failure of \acl{ULTR} and click-based approaches as a whole.

\begin{acks}
We are grateful to Lixin Zou for his valuable insights and clarifications on the original Baidu work, as well as to Jiaxin Mao and Zechun Niu for sharing their learnings from NTCIR 17.

This research was supported by the Mercury Machine Learning Lab, created by TU Delft, the University of Amsterdam, and funded by Booking.com.
Maarten de Rijke was supported by 
the Dutch Research Council (NWO),  project nrs 024.004.022, NWA.1389.20.\-183, and KICH3.LTP.20.006, and the European Union's Horizon Europe program under grant agreement No 101070212.
All content represents the opinion of the authors, which is not necessarily shared or endorsed by their respective employers and/or sponsors.
\end{acks}

\clearpage
\balance
\bibliographystyle{ACM-Reference-Format}
\bibliography{main}

\end{document}